# Multiple magnetic transitions in multiferroic BiMnO$_3$

C. C. Chou,[1] C. L. Huang,[1] S. Mukherjee,[1] Q. Y. Chen,[1] H. Sakurai,[2] A. A. Belik,[2] E. Takayama-Muromachi,[2] and H. D. Yang[1, *]

[1]*Department of Physics, Center for Nanoscience and Nanotechnology, National Sun Yat-Sen University, Kaohsiung 804, Taiwan*
[2]*International Center for Materials Nanoarchitectonics (MANA), National Institute for Materials Science (NIMS), 1-1 Namiki, Tsukuba, Ibaraki 305-0044, Japan*

The magnetic phase variations under hydrostatic pressure on multiferroic BiMnO$_3$ have been examined by the dc magnetization ($M_g(T)$), magnetic hysteresis ($\mu_{eff}(H)$) and ac susceptibility ($\chi'_g(T)$). Three magnetic transitions, manifested as kinks I, II and III on the $M_g(T)$ curves, were identified at 8.7 and 9.4 kbar. With increasing pressure, transition temperatures of kink I and kink II ($T_{kI}$ and $T_{kII}$) tend to decrease, but the temperature of kink III ($T_{kIII}$) showed more complex variation. Under increasing magnetic field, $T_{kI}$ and $T_{kII}$ increase; however, $T_{kIII}$ decreases. Combining $M_g(T)$ curves with $\mu_{eff}(H)$ and $\chi'_g(T)$, more detailed properties of these three kinks would be shown as follows. Kink I is a long-range soft ferromagnetic transition which occurs at $T_{kI} \sim 100$ K under ambient pressure, but is suppressed completely at 11.9 kbar. Kink II emerges at 8.7 kbar along with $T_{kII} \sim 93$ K which is also long-range soft ferromagnetic but canted in nature. Kink III, a canted antiferromagnetic transition, appears at $T_{kIII} \sim 72.5$ K along with kink II also at 8.7 kbar. The proposed phase diagrams at ambient pressure, 9.4 and 11.9 kbar show the different magnetic features of BiMnO$_3$. These findings are believed to result from the variations of crystal structure influenced by the external pressure. These results also indicate the common complicated corelation between the lattice distortion and the spin configuration in multiferroic system.

## I. INTRODUCTION

BiMnO$_3$ undergoes ferroelectric (FE) and ferromagnetic (FM) transitions when cooled below 500 K ($T_E$) and 100 K ($T_C$), respectively.[1,2] In this material, the FM transition is accompanied by a magneto-dielectric anomaly that is characteristic of multiferroicity,[3] such as CdCr$_2$O$_4$,[4] for which FE and FM ordering occur simultaneously. These multiferroics provide a new revenue for novel device applications in many areas of modern technology, including microwave, sensor, transducer, and read/write electronics.[5–7] BiMnO$_3$, as is generally accepted, takes on the monoclinic crystal structure of space group $C2$[1,3,8,9] with off-centered Bi 6s$^2$ lone pairs breaking the antisymmetry.[8] However, Belik *et al*. concluded it with a centrosymmetric $C2/c$.[10] The magnetodielectric anomaly near 100 K, first observed by Kimura *et al*.,[3] was later suggested by Montanari *et al*.[11] as risiing from magnetodielectric and magnetoelastic couplings in $C2/c$ structure.[11] The magnetic properties of BiMnO$_3$ is known to also depend on its oxygen content,[12,13] viz., the oxygen stoichiometric or non-stoichiometry BiMnO$_{3+x}$ in addition to the crystal structure. Belik *et al*. suggested four different structures, namely, monoclinic I (space group : $C2/c$), monoclinic II (space group : $C2/c$), monoclinic P (space group : $P2_1/c$) related to monoclinic II and orthorhombic (space group : $Pnma$).[13] These structures are dictated by the oxygen stoichiometries which are responsible for different transition temperatures. The FM state involves orbital ordering and superexchange interactions.[8–10,14] Heavily distorted MnO$_6$ octahedral in the perovskite compound brings about an orbital-ordering configuration with six superexchange couplings, though not all six Mn-O-Mn bonds are necessarily FM in nature. In fact, four are FM while the other two favors antiferromagnetic (AFM) interactions, manifested effectively in an FM state for T < T$_C$.[1–3,10,11,14–16]

In addition to the chemical stoichiometry and chemically-inflicted distortions, external pressure is also a common means used to effect changes of crystal symmetry through modification of the Mn-O-Mn bond angle and bond length, which then naturally alters the orbital overlapping of the cations and anions.[17,18] The variations of bond angles and bond lengths have a strong effect on the strength of superexchange coupling.[19] Recently, some interesting pressure-dependent studies on ac susceptibility[15] and crystal structures[20] on BiMnO$_3$ were carried out. The result of ac susceptibility reveals a new magnetic state presumably induced by external pressure.[15] Moreover, Belik *et al*. suggested three different phases and two coexistent states could emerged as external pressure increases from ambient pressure to 86 kbar at room temperature (RT).[20]

On a different relevant magnetic material Er$_5$Si$_4$, Magen *et al*. found that external pressure brings about structural variation and correlated magnetic transitions as shown by magnetic hysteresis loops, magnetizations and neutron diffractions.[21] The complex magnetic and structural variations of BiMnO$_3$ emerge from under external pressures.[15,20] As the crystal structure plays an important role in magnetic states,[12,13] the pressure-dependent magnetic properties could well be tied to crystal structure changes influenced by external pressure. In order to obtain a more complete glimpse of such correlations in

BiMnO$_3$, dc magnetization ($M_g(T)$), magnetic hysteresis ($\mu_{eff}(H)$) and ac susceptibility ($\chi'_g(T)$) were measured at various external pressure.

## II. EXPERIMENTAL

BiMnO$_3$ samples were prepared using high purity Bi$_2$O$_3$ and Mn$_2$O$_3$ powders under 60 kbar in a belt-type high-pressure apparatus at 1383 K for 60-70 minutes as described in the literature.[10,13,16,20] RT X-ray diffraction was performed with a diffractometer (Model : D5000, Siemens) by using CuK$\alpha$ as the radiation source in a 2$\theta$ range from 20° to 70°. The XRD pattern (the inset of Fig. 1 (a)) has clearly revealed single phase monoclinic structure which is identical to earlier reports.[1,3] The magnetization were measured using a superconducting quantum interference device (SQUID) magnetometer (Model : MPMS-XL7, Quantum Design) with temperature varying from 2 to 300 K at different magnetic fields and pressures both in zero-field-cooled (ZFC) and field-cooled (FC) modes. The magnetic hysteresis loops were measured within the range of ±70 kOe. Frequency-dependent ac susceptibility measurements were carried out at ambient pressure and 11.9 kbar. For the sake of comparison, the hydrostatic-pressure effects on $M_g(T)$, $\mu_{eff}(H)$ and $\chi'_g(T)$ were investigated up to 11.9 kbar using the piston-cylinder self-clamped technique.[18,21] An inert fluid, Daphne-7373,[22] was used as a pressure transmitting medium along with a manometer made of superconducting tin.

## III. RESULTS AND DISCUSSION

### A. pressure-dependent magnetization at 50 Oe

Complex behaviors of three kinks are observed in the $M_g(T)$ curves (FC mode) with 50 Oe under various pressure ($p_{max} \sim 11.9$ kbar), as shown in Fig. 1 (a). All of the kink temperatures, i.e., $T_{kI}$, $T_{kII}$ and $T_{kIII}$, whose existence reflects a transition of some sort, are decided by the derivative approach based on d$M_g$/d$T$ minima of the $M_g(T)$ curve. An illustration of this process is given in the inset of Fig. 1 (b). The pressure-dependent phase diagram according to the kink temperatures, as shown in Fig. 1 (b). The salient features at various pressures are as follows. (1) $T_{kI}$ decreases with increasing pressure, and could not be noticed at 11.9 kbar. In the inset of Fig. 1 (b), the original FM transition (kink I), explicit at 100 K under ambient pressure, is suppressed by increasing pressure, and completely disappears at 11.9 kbar; (2) $T_{kII}$ dwindles along with $T_{kI}$ as a result of increasing pressure; (3) $T_{kIII}$ is 72.5 K at 8.7 kbar, while increases at 9.4 kbar, but decreases at 11.9 kbar. The variation implies the wax and wane of some magnetic ordering; (4) Kink II and kink III simultaneously emerge at 8.7 kbar, and still subsist at 11.9 kbar. It seems that kink II and kink

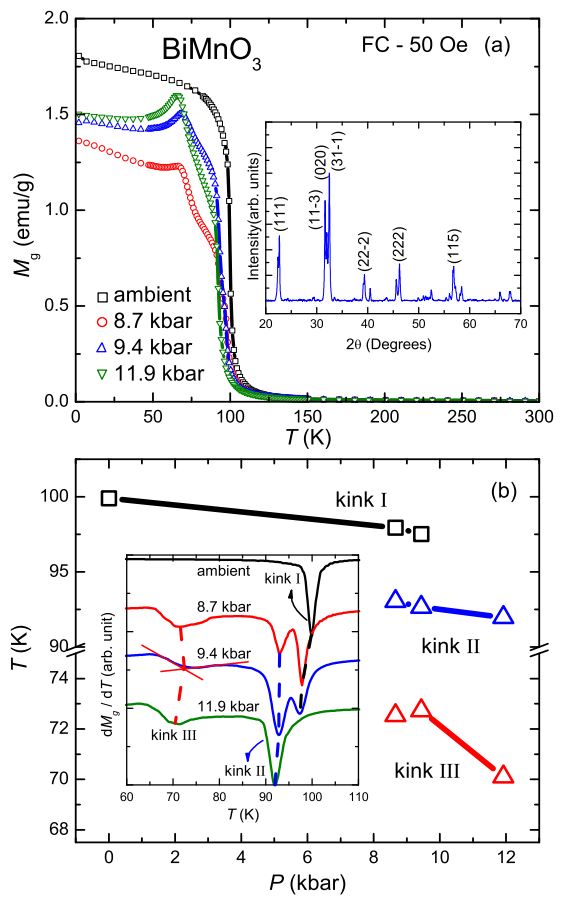

FIG. 1: (Color online) (a) $M_g(T)$ curves with different pressures in 50 Oe. The inset of (a) is the normalized powder X-ray diffraction pattern between 20° to 70°. (b) The pressure dependence of kink temperatures (FC mode), which is derived from the inset of (b) showing the d$M_g$/d$T$ curves under different pressures.

III indicate the newly stable magnetic transitions under high pressure; (5) The pressures, 8.7 and 9.4 kbar, where three kinks simultaneously are observed, agree with our earlier work[15] and the phase diagram reported by Belik et al.[20] (6) In Fig. 1 (a), the phenomenon that kink I and kink II have a sharp rise in $M_g(T)$ curve is an evidence of the FM state while the decline of magnetization below $T_{kIII}$ implies a re-entrant AFM transition taking place at this temperature. The steeper falls of kink III at 9.4 and 11.9 kbar suggest a more drastic change into the AFM state.

### B. magnetic-field dependent magnetization and magnetic hysteresis loop at ambient pressure, 9.4 and 11.9 kbar

In order to understand the nature of these phases appearing at high pressure, the $\mu_{eff}(H)$ and magnetic-field dependent $M_g(T)$ curves under ambient pressure, 9.4 and

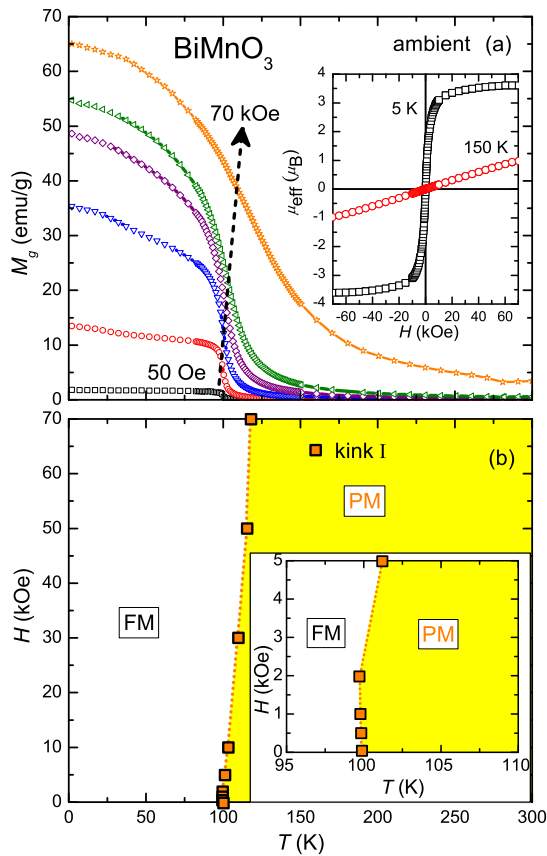

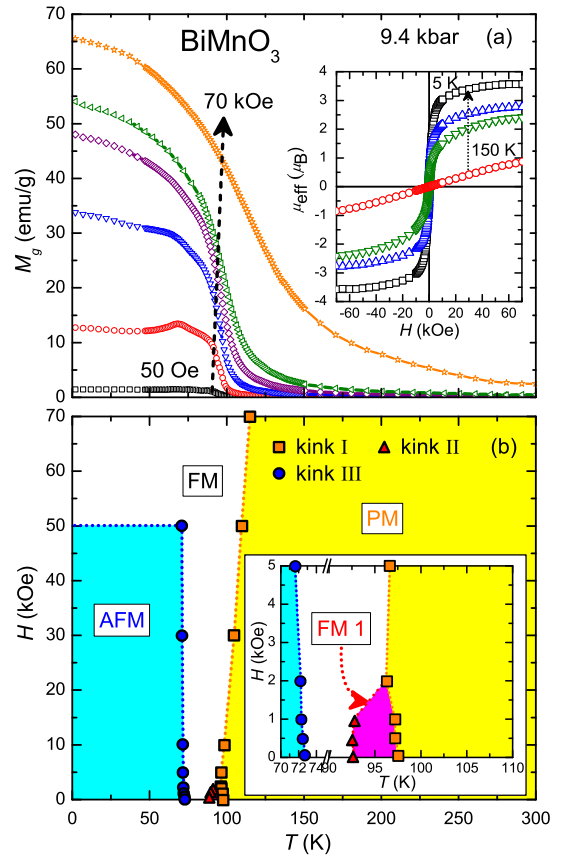

FIG. 2: (Color online) (a) $M_g(T)$ curves with different magnetic fields (0.05, 0.5, 2, 5, 10 and 70 kOe) at ambient pressure. The inset of (a) is the $\mu_{eff}(H)$ hysteresis loop at 5 and 150 K. (b) The proposed temperature-dependent magnetic states with different magnetic fields (0.05, 0.5, 1, 2, 5, 10, 30, 50 and 70 kOe) at ambient pressure. The inset of (b) is the expanded part at small magnetic fields.

FIG. 3: (Color online) (a) $M_g(T)$ curves with different magnetic fields (the same conditions in Fig. 2) at 9.4 kbar. The inset of (a) is the $\mu_{eff}(H)$ hysteresis loop at 5, 80, 95 and 150 K. (b) The proposed temperature-dependent magnetic states with different magnetic fields at 9.4 kbar. The inset of (b) is the expanded part between $T_{kI}$ and $T_{kII}$.

11.9 kbar were measured. The $M_g(T)$ curves at ambient pressure are represented in Fig. 2 (a). With increasing magnetic field, kink I shifts to a higher temperature, indicating enhanced spin alignment. This behavior is consistent with the modified Ising model developed by Hassink et al.[23] The inset of Fig. 2 (a) shows the $\mu_{eff}(H)$ curves at 5 and 150 K. At 5 K, the fast saturated $\mu_{eff}(H)$ at $H > 10$ kOe, and the small yet noticed loop are the convincing evidence of soft ferromagnetism. Nevertheless, the linear $\mu_{eff}(H)$ curve at 150 K suggests a paramagnetic (PM) state. All of the above observed features can be summarized into an ambient-pressure phase diagram, shown in Fig. 2 (b), with two clearly delineated magnetic states with $T_{kI}$. The $T_{kI}(H)$ is the demarcation line between the PM and FM states. As illustrated the inset of Fig. 2 (b), $T_{kI}$ decreases first, but increases when $H > 2$ kOe. This intriguing behavior is also observed at 9.4 and 11.9 kbar, shown in the insets of Figs. 3 (b) and 4 (b), respectively; however, this phenomena is not yet clearly understood.

At 9.4 kbar, the magnetic-field dependent $M_g(T)$ (Fig. 3 (a)) shows some distinctively different behaviors. (1) $T_{kI}$ increases with increasing magnetic field; (2) $T_{kII}$ also increases with increasing magnetic field, but is convolute with kink I for $H > 5$ kOe; (3) Kink I and kink II define the FM transition with sharply-rising $M_g$ value ; (4) $T_{kIII}$ blurs with increasing magnetic field, and can not be traced beyond 70 kOe; (5) The fall of magnetization for $T < T_{kIII}$ implies that $T_{kIII}$ is probably an AFM transition temperature. The $\mu_{eff}(H)$ curve shown in the inset of Fig. 3 (a) also exhibits a small but noticeable hysteresis loop at 5 K. The FM ordering has formed at higher temperatures by the curves at 85 and 95 K while the loop at 150 K points to a PM scenario. Considering these results, there are four magnetic regions for BiMnO$_3$ at 9.4 kbar, shown in Fig. 3 (b). (1) Above $T_{kI}$, PM state exists in high temperature; (2) For $H < 5$ kOe, the magnetic state between $T_{kI}$ and $T_{kII}$ seems to be FM 1; (3) The FM state is observed below $T_{kI}$ ($H > 2$ kOe) and $T_{kII}$ ($H < 2$ kOe); (4) The AFM state appears in low temperature ($T < T_{kIII}$).

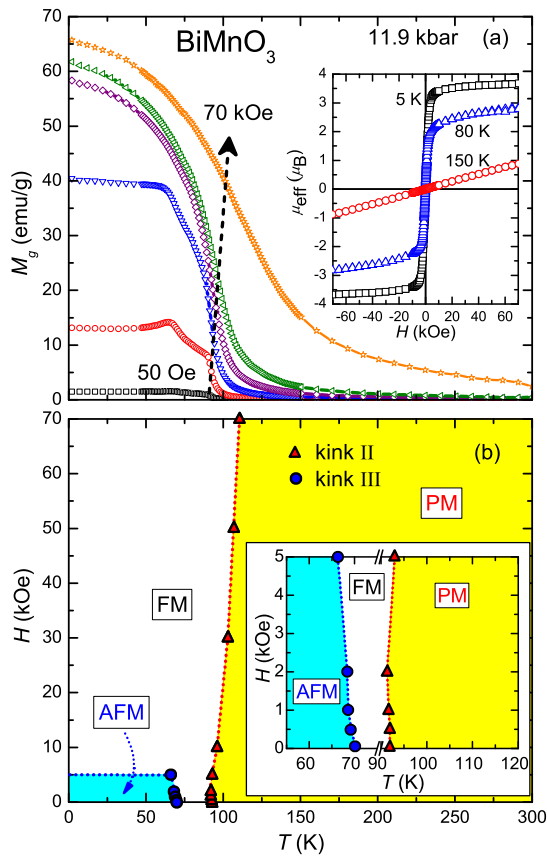

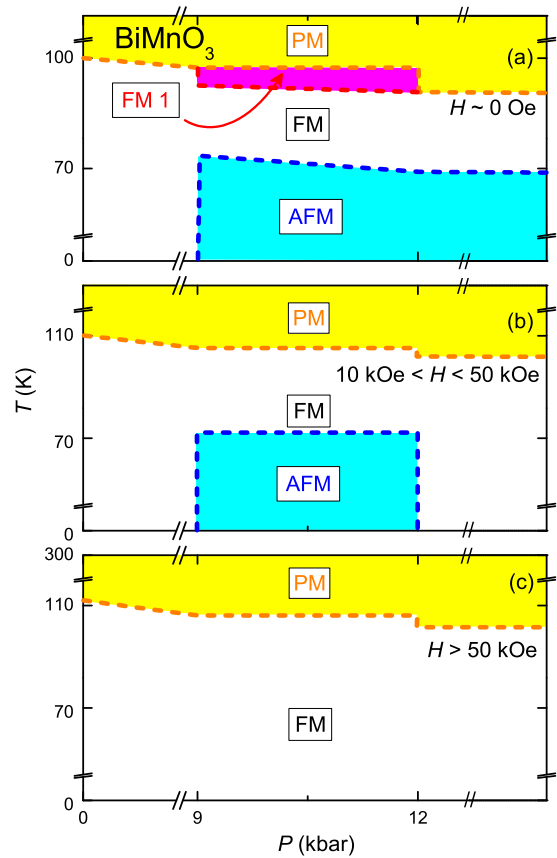

FIG. 4: (Color online) (a) $M_g(T)$ curves with different magnetic fields (the same conditions in Fig. 2) at 11.9 kbar. The inset of (a) is the $\mu_{eff}(H)$ hysteresis loop at 5, 80 and 150 K. (b) The proposed temperature-dependent magnetic states with different magnetic fields at 11.9 kbar. The inset of (b) is the expanded part at small magnetic fields.

FIG. 5: (Color online) The $T$-$P$ magnetic phase diagram at (a) $H \sim 0$ Oe, (b) 10 kOe $< H <$ 50 kOe and (c) $H >$ 50 kOe

At the highest pressure (11.9 kbar), the magnetic-field dependent $M_g(T)$ curves (Fig. 4 (a)) exhibit following features. (1) Kink I disappears; however, both kink II and kink III still exist; (2) $T_{kII}$ goes higher with increasing magnetic field; (3) $T_{kIII}$ decreases with increasing magnetic field, but the variation is not recognizable based on $dM_g/dT$ for $H >$ 10 kOe, suggestive of a weaker AFM orderings at 11.9 kbar as compared to that at 9.4 kbar; (4) Below $T_{kIII}$, the drop of magnetization implies emergence of AFM state. The $\mu_{eff}(H)$ curves were measured at 5, 80 and 150 K, shown in the inset of Fig. 4 (a). The PM property is observed at 150 K, and the FM ordering exists at 5 and 80 K. At 5 K, the small $\mu_{eff}(H)$ hysteresis loop still could be noticed at this pressure. According to these observations, a suggested phase diagram of the magnetic states is shown in Fig. 4 (b), displaying three separate magnetic states staying for the BiMnO$_3$ at 11.9 kbar. Note that (1) the PM state is observed above $T_{kII}$, (2) the FM state emerges below $T_{kII}$ and (3) at $T < T_{kIII}$, it enters into the AFM state. The above-mentioned discussions are summarized as sketched in the

$T$-$P$ magnetic phase diagram in different magnetic-field regions, as shown in Fig. 5.

The $\mu_{eff}(H)$ data at 5 K under different pressures are worthy of a detailed discussion. All of these curves show sharp variations between $\pm$10 kOe, and go to near complete saturation over this range. The coercive field ($H_c$) and the remnant magnetization ($M_r$) are varied from 12.29 to 37.74 Oe and from 16.97 to 62.88 $\times 10^{-3}\mu_B$, respectively. Such small values imply a soft FM property on BiMnO$_3$. In addition, the saturation magnetization ($M_s$) falls within 3.57 to 3.67 $\mu_B$, an evidence that the valence of Mn is not changed by pressure even if the crystal structure has changed.[20]. In order to further probe the magnetic effects, we have conducted $\chi'_g(T)$ measurement.

### C. frequency-dependent ac susceptibility at ambient pressure and 11.9 kbar

The frequency-dependent $\chi'_g(T)$ curves at ambient pressure and 11.9 kbar are depicted in Fig. 6 (a). The FM transition at ambient pressure remains at 98 K for all frequencies, but an anomaly is observed below 90 K,

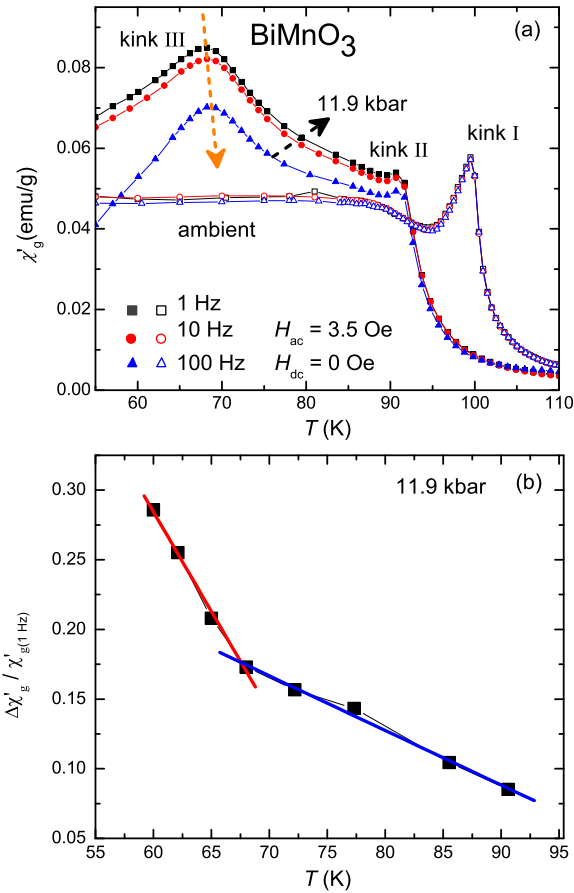

FIG. 6: (Color online) (a) The frequency-dependent $\chi'_g(T)$ at ambient pressure and 11.9 kbar with $H_{ac} = 3.5$ Oe and $H_{dc} = 0$ Oe. The full and empty symbols are the data at ambient pressure and 11.9 kbar, respectively. (b) The temperature-dependent $\Delta \chi'_g/\chi'_{g(1\ Hz)}$ curve. ($\Delta\chi'_g = |\chi'_{g(100\ Hz)} - \chi'_{g(1\ Hz)}|$)

a phenomenon referred to by Belik et al.[16,24] as a spin-glass-like state. Here, $T_{kII}$ and $T_{kIII}$ are 90 and 68 K under 11.9 kbar, respectively. Regarding the variation of temperature, frequency-independent phenomena of $T_{kI}$ and $T_{kII}$ at low frequencies point to a long-range FM ordering.[25] However, the relative shift of $T_{kIII}$ corresponding to peak value of $\chi'_g$ per decade of frequency, i.e., $\Delta T_f/[T_f\Delta(\log \omega)]$ is determined to be $5.88\times10^{-3}$ at 1 and 100 Hz. This value is in the range of that ($\sim 10^{-3}$) obtained in canonical spin-glass.[25] The spin-glass ordering appears along with the AFM transition exhibiting a coupling between each other. In respect of intensities of the kinks, kink I almost remains the same value, a manifestation of a short relaxation. However, the intensities of kink II and kink III unfold large variations, an indication of a long relaxation time. This phenomenon implies that kink II and kink III probably are canted FM and canted AFM orderings, respectively. In the Fig. 6 (b), the slope of the temperature-dependent $\Delta \chi'_g/\chi'_{g(1\ Hz)}$ curve ($\Delta\chi'_g = |\chi'_{g(100\ Hz)} - \chi'_{g(1\ Hz)}|$) shows a visible change below $T_{kIII}$. The distinctive variation of slope implies a signature that two different magnetic ordered states exist.

The $M_g(T)$ and $\chi'_g(T)$ curves show new magnetic transitions (kink II and kink III) at high pressure while the original FM ordering could not be observed at the highest pressure 11.9 kbar. Pressure may change the valence state of Mn; however, the similar values of $M_s$ at 5 K under different pressures reveal new magnetic states are not coming from the various valences. Because the pressures of coexistent kinks in our investigation are 8.7 and 9.4 kbar which agree with the recent studies about ac susceptibility[15] and crystal structure[20] under high pressure, the complex magnetic transitions of $BiMnO_3$ probably come from the structural variations from the monoclinic structure $C2/c$ to $P2_1/c$ induced by pressure.[20]

### D. differences and similarities between multiferroics $BiMnO_3$ and $YMnO_3$

Among $RMnO_3$ manganites[26] (R = Ho, Er, Tm, Yb, Lu, Y and Sc), which have been extensively studied to show the multiferroic behavior,[27–29] for implication, we take the $YMnO_3$ as an example to compare differences and similarities with $BiMnO_3$. (1) $T_{kI}$ of $BiMnO_3$ is decreased, and suppressed completely at high pressure; on the contrary, $T_N$ of $YMnO_3$ is increased with increasing pressure ($dT_N/dP = 0.29$ K/kbar).[30] (2) Changing the manganese or oxygen content of sample brings about the variation of spin configuration and/or lattice structure.[12,13,31] (3) Under pressure, new magnetic phases, namely, new magnetic transitions (kink II and kink III) in $BiMnO_3$ and a pressure-induced spin-liquid phase of $YMnO_3$.[32] were observed in both systems. Interestingly, by contrast, the non-multiferroic manganite $La_{1-x}Sr_xMnO_3$ reveals positive $dT_C/dP$ but no new phase induced by pressure on a lattice Mn or O off-stoichiometry.[33] (4) The complex interaction between the lattice distortion and the spin configuration seems a common phenomenon in multiferroic system, induced by changing the Mn or O content of samples or adding external pressure. These mentioned differences and similarities also remind that the application using this class of materials should take into account this complexity.

### IV. CONCLUSION

Pressure-dependent $M_g(T)$, $\mu_{eff}(H)$ and $\chi'_g(T)$ curves of multiferroic $BiMnO_3$ are measured, showing various magnetic transitions. (1) $T_{kI}$ (100 K under ambient pressure) is long-range soft FM transition temperature which increases with increasing magnetic field, while deceases under high pressure, and could not be observed at 11.9 kbar completely; (2) Kink II as well as kink I is a long-range soft FM but canted transition, which has the similar variations to kink I under various magnetic field and pressure. Nevertheless, $T_{kII}$ (93 K under 8.7 kbar) emerges at 8.7 kbar, and still could be noticed at 11.9 kbar. (3) Kink III is a canted AFM transition with frequency-dependent variation under high pressure. $T_{kIII}$ existing along with $T_{kII}$ (72.5 K under 8.7 kbar) from 8.7 to 11.9 kbar decreases with increasing magnetic field, has the wax and wane of magnetic ordering with increasing pressure. Besides, the proposed magnetic-field dependent phase diagrams at ambient pressure, 9.4 and 11.9 kbar are shown. The valence of Mn is the same under different pressures. We suggest that all of these phenomena should be caused by the variation from the monoclinic structure $C2/c$ to $P2_1/c$.[20] The results imply the common complicated corelation between the lattice distortion and the spin configuration exists in multiferroic system.

## V. ACKNOWLEDGMENTS

This work was supported by National Science Council of Taiwan under Grant No. NSC 97-2112-M-110-005-MY3, World Premier International Research Center Initiative (WPI Initiative, MEXT, Japan) and the NIMS Individual-Type Competitive Research Grant.